\documentclass[sigconf]{acmart}
\usepackage{booktabs}
\usepackage{graphicx}
\usepackage{amsmath}

\newcommand{\gfBestParaAll}{+15.5}
\newcommand{\gfBestParaAllTests}{5}

\newcommand{\gfBestParaThird}{+25.7}

\newcommand{\gfBetween}{13}
\newcommand{\gfCitedPerQ}{3}
\newcommand{\gfCitedRetrievedPct}{100.0}
\newcommand{\gfCitedURLs}{22,315}
\newcommand{\gfCollectedDayToDay}{65}
\newcommand{\gfConsultedPerQ}{26}
\newcommand{\gfDeepGemCitGoogleAbsent}{93}

\newcommand{\gfDeepGptRetBingAbsent}{99}

\newcommand{\gfDeepGptRetGoogleAbsent}{97}
\newcommand{\gfDeepGptRetGoogleSiteFound}{51}
\newcommand{\gfFaqPooled}{+18.1}
\newcommand{\gfFaqWithin}{\textminus3.4}
\newcommand{\gfGemAddBrand}{5}
\newcommand{\gfGemAddYear}{2}

\newcommand{\gfGemMentionThirdOR}{6.3}
\newcommand{\gfGemQueriesPerAnswer}{3}
\newcommand{\gfGemSimilarity}{0.78}

\newcommand{\gfGptAddBrand}{28}
\newcommand{\gfGptAddYear}{23}

\newcommand{\gfGptMentionNone}{46}

\newcommand{\gfGptMentionPriorOR}{5.8}

\newcommand{\gfGptMentionThirdOR}{4.2}
\newcommand{\gfGptQueriesPerAnswer}{4}
\newcommand{\gfGptSimilarity}{0.66}
\newcommand{\gfGptWordsKept}{43}

\newcommand{\gfMultiCallOR}{10.8}

\newcommand{\gfPageLinks}{+22.7}

\newcommand{\gfPageVar}{2.8}

\newcommand{\gfPriorAddIfIn}{67}
\newcommand{\gfPriorAddIfOut}{11}
\newcommand{\gfPriorBaseline}{25}
\newcommand{\gfPriorCitations}{47}
\newcommand{\gfPriorRankFactor}{1.38}
\newcommand{\gfPriorSelOR}{1.60}
\newcommand{\gfRankElevenTwenty}{3.5}

\newcommand{\gfRankOne}{43.8}
\newcommand{\gfRankOneHi}{45.1}
\newcommand{\gfRankOneLo}{42.6}
\newcommand{\gfRankSixTen}{12.2}

\newcommand{\gfRankTwo}{35.5}

\newcommand{\gfReplayDays}{11}
\newcommand{\gfReplayInCollected}{59}
\newcommand{\gfReplayLastDay}{7 October 2026}
\newcommand{\gfReplayRuns}{13}
\newcommand{\gfRetrieval}{32}
\newcommand{\gfRetrievalHi}{42}
\newcommand{\gfRetrievalLo}{24}
\newcommand{\gfRetrievalMinusPage}{29}
\newcommand{\gfRetrievalMinusPageHi}{38}
\newcommand{\gfRetrievalMinusPageLo}{20}
\newcommand{\gfRhoKeywords}{0.08}
\newcommand{\gfRhoMeaning}{0.20}
\newcommand{\gfSameCited}{15}

\newcommand{\gfSameQueries}{2.7}

\newcommand{\gfSelBrand}{1.71}
\newcommand{\gfSelComparison}{0.34}
\newcommand{\gfSelHeadings}{1.28}
\newcommand{\gfSelLength}{0.59}
\newcommand{\gfSelList}{0.54}

\newcommand{\gfSelPage}{32.5}

\newcommand{\gfSelRank}{26.8}

\newcommand{\gfSelReddit}{0.30}
\newcommand{\gfShuffled}{0.05}
\newcommand{\gfShuffledP}{0.005}
\newcommand{\gfSiteAuthority}{+12.1}
\newcommand{\gfSourceLists}{4946}

\newcommand{\gfTitleAll}{+14.1}
\newcommand{\gfTitleAllTests}{5}
\newcommand{\gfTitleBrand}{+7.4}
\newcommand{\gfTitleBrandTests}{4}
\newcommand{\gfTitleThird}{+23.6}

\newcommand{\gfVarPages}{6178}
\newcommand{\gfVarSites}{362}
\newcommand{\gfWindowFirst}{14}
\newcommand{\gfWindowLast}{42}

\newcommand{\gfZeroIfNotSearched}{83.1}
\newcommand{\gfZeroIfSearched}{7.1}

\newcommand{\woBestParaAll}{+12.9}
\newcommand{\woBestParaAllTests}{5}

\newcommand{\woBetween}{9}

\newcommand{\woCitedRetrievedPct}{100.0}
\newcommand{\woCitedURLs}{2,107}

\newcommand{\woDeepGptRetGoogleAbsent}{99}

\newcommand{\woFaqPooled}{+22.5}
\newcommand{\woFaqWithin}{\textminus1.9}

\newcommand{\woGemMentionThirdOR}{17.7}

\newcommand{\woGptAddBrand}{3}

\newcommand{\woGptMentionThirdOR}{39.7}

\newcommand{\woPageLinks}{+16.3}

\newcommand{\woPageVar}{10.0}

\newcommand{\woPriorCitations}{32}

\newcommand{\woRankElevenTwenty}{8.3}

\newcommand{\woRankOne}{57.1}

\newcommand{\woRetrieval}{30}
\newcommand{\woRetrievalHi}{43}
\newcommand{\woRetrievalLo}{15}

\newcommand{\woRetrievalMinusPageHi}{31}
\newcommand{\woRetrievalMinusPageLo}{\textminus1}

\newcommand{\woSiteAuthority}{\textminus1.3}

\newcommand{\woTitleAll}{+26.4}
\newcommand{\woTitleAllTests}{5}
\newcommand{\woTitleBrand}{+33.6}
\newcommand{\woTitleBrandTests}{5}

\newcommand{\woVarPages}{629}

\newcommand{\AppTabQuestions}{%
\begin{tabular}{@{}lr@{\hspace{1.2em}}lr@{}}
\toprule
\multicolumn{2}{@{}l}{Consumer (651)} & \multicolumn{2}{l@{}}{Niche (55)} \\
\midrule
Customer Support & 100 & AEC Industry Focus & 10 \\
Equipment & 100 & AI Capabilities & 10 \\
Competitive Landscape & 100 & Knowledge Management & 10 \\
Speed & 100 & Proposal Automation & 10 \\
Pricing & 100 & Workflow Integration & 10 \\
Performance & 98 & Competitive Landscape & 5 \\
Reliability & 53 &  &  \\
\midrule
\emph{learning} & 298 & \emph{learning} & 52 \\
\emph{comparing} & 264 & \emph{comparing} & 3 \\
\emph{ready to buy} & 89 &  &  \\
\bottomrule
\end{tabular}}
\newcommand{\AppTabTests}{%
\begin{tabular}{@{}lrccccc@{\hspace{0.8em}}rc@{}}
\toprule
 & \multicolumn{6}{c}{Consumer market} & \multicolumn{2}{c@{}}{Niche market} \\
\cmidrule(lr){2-7}\cmidrule(l){8-9}
Feature & Effect & 1 & 2 & 3 & 4 & 5 & Effect & Tests \\
\midrule
\textbf{Title match} & +14.1 & $\bullet$ & $\bullet$ & $\bullet$ & $\bullet$ & $\bullet$ & +26.4 & \textbf{5} \\
\textbf{Best-paragraph match} & +15.5 & $\bullet$ & $\bullet$ & $\bullet$ & $\bullet$ & $\bullet$ & +12.9 & \textbf{5} \\
Word overlap & +9.1 & $\bullet$ & $\circ$ & $\bullet$ & $\bullet$ & $\bullet$ & +13.0 & 4 \\
First-paragraph match & +5.9 & $\bullet$ & $\circ$ & $\bullet$ & $\bullet$ & $\bullet$ & +0.1 & 0 \\
Length (words) & +0.9 & $\circ$ & $\circ$ & $\circ$ & $\circ$ & $\bullet$ & +2.2 & 1 \\
Outbound links & +1.4 & $\circ$ & $\circ$ & $\circ$ & $\circ$ & $\bullet$ & +3.2 & 1 \\
Headings & \textminus{}2.7 & $\circ$ & $\circ$ & $\circ$ & $\bullet$ & $\bullet$ & +7.7 & 4 \\
Tables & \textminus{}1.4 & $\circ$ & $\circ$ & $\circ$ & $\circ$ & $\bullet$ & \textminus{}2.5 & 1 \\
Lists & \textminus{}1.0 & $\circ$ & $\circ$ & $\circ$ & $\circ$ & $\bullet$ & \textminus{}4.8 & 1 \\
FAQ block$^\dagger$ & \textminus{}3.4 & $\circ$ & $\circ$ & $\circ$ & $\bullet$ & $\bullet$ & \textminus{}1.9 & 2 \\
Summary box$^\dagger$ & +6.9 & $\circ$ & $\circ$ & $\circ$ & $\bullet$ & $\bullet$ & +8.0 & 1 \\
Author$^\dagger$ & \textminus{}4.6 & $\circ$ & $\circ$ & $\circ$ & $\bullet$ & $\bullet$ & +2.0 & 1 \\
Schema.org markup$^\dagger$ & \textminus{}5.3 & $\circ$ & $\circ$ & $\circ$ & $\bullet$ & $\bullet$ & +16.2 & 2 \\
Page age & \textminus{}5.1 & $\bullet$ & $\circ$ & $\bullet$ & $\bullet$ & $\bullet$ & \textminus{}2.8 & 1 \\
\bottomrule
\end{tabular}}
\newcommand{\AppTabEngines}{%
\begin{tabular}{@{}lrrr@{}}
\toprule
Feature & ChatGPT & Gemini & Perplexity \\
\midrule
Title match & \textminus{}0.2 & +16.8$^{*}$ & +4.6$^{*}$ \\
Best-paragraph match & +9.5$^{*}$ & +7.8$^{*}$ & +4.8$^{*}$ \\
Word overlap & +2.4 & +5.3$^{*}$ & +7.1$^{*}$ \\
First-paragraph match & +0.7 & +8.6$^{*}$ & +0.2 \\
FAQ block$^\dagger$ & \textminus{}1.7 & \textminus{}0.3 & \textminus{}2.7 \\
Schema.org markup$^\dagger$ & \textminus{}0.0 & +1.5 & \textminus{}0.5 \\
Page age & \textminus{}4.2$^{*}$ & \textminus{}5.2$^{*}$ & +1.9 \\
\bottomrule
\end{tabular}}
\newcommand{\MainTabEffectsRows}{%
Best paragraph matches & \textbf{+15.5 (5)} & +5.2 (2) & \textbf{+25.7 (5)} \\
Title matches a query & \textbf{+14.1 (5)} & +7.4 (4) & \textbf{+23.6 (5)} \\
Word overlap with queries & +9.1 (4) & +11.9 (3) & \textbf{+12.8 (5)} \\
First paragraph matches & +5.9 (4) & +7.8 (4) & +3.1 (3) \\
Page age & \textminus{}5.1 (4) & \textminus{}10.6 (3) & \textminus{}4.2 (3) \\
Word count & +0.9 (1) & +2.9 (1) & \textminus{}3.0 (2) \\
Outbound links & +1.4 (1) & +6.6 (1) & +0.5 (0) \\
Headings & \textminus{}2.7 (2) & \textminus{}1.4 (1) & +1.9 (1) \\
Tables & \textminus{}1.4 (1) & +12.5 (3) & \textminus{}4.0 (1) \\
Lists & \textminus{}1.0 (1) & \textminus{}18.1 (3) & +0.5 (1) \\
Has FAQ block & \textminus{}3.4 (2) & \textminus{}1.9 (2) & \textminus{}9.2 (3) \\
Has summary box & +6.9 (2) & +2.6 (2) & +5.1 (2) \\
Has author & \textminus{}4.6 (2) & \textminus{}15.8 (4) & +3.7 (2) \\
Has schema.org markup & \textminus{}5.3 (2) & \textminus{}12.2 (2) & +11.7 (2) \\
\midrule
Pages / websites & 12,038 / 1,120 & 4,794 / 77 & 7,185 / 1,044 \\}

\newcommand{\pending}[1]{\textbf{[#1]}}
\graphicspath{{../figures/paper/}{../figures/}}

\setcopyright{none}

\begin{document}

\title{What Drives AI Citations? Content, Retrieval and Model Knowledge in Generative Engine Optimization (GEO)}

\author{Mahe Chen}
\email{mahe.chen@mail.utoronto.ca}
\affiliation{\institution{University of Toronto}\country{Canada}}

\author{Nick Koudas}
\email{koudas@cs.toronto.edu}
\affiliation{\institution{University of Toronto}\country{Canada}}

\begin{abstract}
Generative search engines such as ChatGPT, Gemini and Perplexity now decide which Web pages users see as
sources, and a growing industry of generative engine optimization advises publishers on how to earn these
citations. We study what shapes citation in two 90-day collections of answers to commercial questions in
contrasting markets: 158,099 answers with 468,037 citations. Comparing pages of the same website, content
matters in one way: pages whose title and paragraphs match the questions people ask are cited more, whereas
FAQ blocks, structured data, summary boxes and page speed appear effective only across websites. Page
properties nevertheless explain little of the variation between pages of a website. Replaying the questions
through the engines' APIs shows why. Every page ChatGPT cites is one its own search retrieved, and which
pages that search retrieves, for sub-queries the model rewrites on every answer and that rarely coincide
with Google's or Bing's results, explains far more of the variation than page properties (\gfRetrieval\%
against \gfPageVar\%). Content acts mainly through this step, by matching the engine's queries in meaning.
The model's own knowledge shapes it too: ChatGPT writes brands it knows into its searches, prefers websites
it knows, and names brands without any supporting page, and controlled experiments show that brand names
reorder its rankings of products with identical facts. Pages are thus cited when searched; brands are also
named when remembered. The findings replicate in a market of small B2B vendors. We release data and code.
\end{abstract}



\maketitle

\section{Introduction}\label{sec:intro}

Generative search engines such as ChatGPT, Gemini and Perplexity answer a question with a written text
and a handful of links. For many users, these links have become the entry point to the Web, and the
engines thereby decide which Web pages, and which brands, are seen. A fast-growing industry of
\emph{generative engine optimization} (GEO) advises publishers on how to earn these citations, most often by
adding FAQ blocks, structured data, summary boxes and author information to their pages. Whether such
advice works is largely untested outside controlled settings, and observational comparisons are easily
misled: practices are concentrated on popular websites, which are cited more for reasons that have nothing
to do with the practice.

In this paper, we study what shapes citation in generative search engines, using two 90-day collections of
daily answers to commercial questions in contrasting markets: a consumer market dominated by a few large
brands (US home internet; 651 questions, 143,249 answers, 396,188 citations) and a niche market of small
business-software vendors (55 questions, 14,850 answers, 71,849 citations). We address three research
questions. \textbf{RQ1:} which properties of a page's content go with more citations, when pages are
compared within the same website? \textbf{RQ2:} how much of the variation in citations do page properties
explain, and what is the rest? \textbf{RQ3:} how do the engine's retrieval and the model's own knowledge
shape which pages are cited and which brands are named?

To answer them, we combine four approaches. A within-website design with a five-test robustness rule
separates the effect of a page from that of its website. An API replay of every question exposes what the
collection cannot show: the searches each engine writes, the ranked list of pages it retrieves, and which
of them it cites. A probe of the engines' models without search measures what they already know. Finally,
controlled experiments vary one factor at a time in the input the model reads: the order of the sources,
the names of brands and publishers, and the set of alternatives.

The results can be summarized in the title's three terms. \emph{Content} matters in one way: pages whose
title and paragraphs match the questions people ask are cited more, whereas FAQ blocks, structured data,
summary boxes, author boxes and page speed appear effective only across websites. Page properties,
however, explain little of the variation between pages of a website (\gfPageVar\% for ChatGPT), although
citation counts are highly repeatable. The reason lies in \emph{retrieval}: every page ChatGPT cites is
one its own search retrieved, and retrieval explains \gfRetrieval\% of the same variation. The engines
rewrite each question into new sub-queries on every answer, rank pages by how well they match these
queries in meaning, and cite mostly from the top of their own ranking, which is not that of Google or Bing.
Finally, the \emph{model's knowledge} shapes both steps and the answer itself: ChatGPT writes brands it knows
into its searches, prefers websites it knows, and names brands without any supporting page
(\gfGptMentionNone\% of its mentions), and in controlled experiments brand names reorder its rankings of
products with identical facts. Pages are thus cited when they are searched; brands are also named when they
are remembered. A second market of small vendors replicates these findings and shows where they depend on
the market: there, the models barely know the brands, and a stable position exists only when search finds
evidence about them.

Our contributions are as follows:
\begin{itemize}
\item Two longitudinal datasets of generative search answers in contrasting markets, joined with page
content, links and replay logs, released with all code (Section~\ref{sec:data}).
\item A within-website protocol for citation studies, under which the recommended practices lose their
apparent effect while matching the questions survives five tests (Section~\ref{sec:rq1}).
\item A decomposition of citations into website, page, chance and drift, showing that most of the
variation is repeatable but unexplained by page properties and changes over weeks (Section~\ref{sec:rq2}).
\item An API replay method that makes retrieval observable, with evidence that citation follows the
engine's own ranking, which matches rewritten queries in meaning and almost never returns the pages Google
or Bing return, even for Gemini (Section~\ref{sec:rq3}).
\item Evidence that the model's prior enters retrieval, selection and the answer, and controlled
experiments showing that source order and brand names change its choices (Sections~\ref{sec:prior}
and~\ref{sec:answerstep}).
\end{itemize}
As generative engines become a gateway to the Web, which pages and brands are visible depends on retrieval
and model knowledge that publishers cannot observe and that change from week to week.

\section{Related Work}\label{sec:related}

\paragraph{Generative search engines and their citations.} Early work evaluated whether the citations of
generative search engines support their statements, and found that many do not~\cite{liu2023verifiability}.
Measurement studies have since described which sources the engines cite: audits found commercial and
geographic biases in the sources of answers on public topics~\cite{li2024arbiters}, citations of news that
concentrate on few outlets~\cite{yang2025news}, a strong preference for third-party (earned) media over
brand-owned content~\cite{chen2025geo}, and sources that differ substantially between Google Search, its
AI Overviews and Gemini~\cite{grossman2026disrupts}\cite{maheNavig}. Others related page characteristics to how strongly a
cited page is used in the answer~\cite{zhang2026absorption}, and showed that visibility varies across
repeated queries and must be measured as a distribution~\cite{schulte2026measure}. These studies observe
the answers. In contrast, we hold the website fixed when comparing pages, follow the same questions daily
for 90 days, and replay them through the engines' APIs to observe the retrieval step itself, the searches,
the ranked pages and the selection among them, which the answers do not reveal.

\paragraph{Optimizing content for generative engines.} Generative engine optimization was introduced as
rewriting a page so that it is more visible in answers generated from a given set of
sources~\cite{aggarwal2024geo}. Related work showed that injected text can promote a product in the
rankings of conversational search engines~\cite{pfrommer2024ranking} and of LLM
recommendations~\cite{kumar2024manipulating}, and a recent benchmark found that most such methods are
ineffective or harmful, whereas conventional search engine optimization remains
effective~\cite{puerto2025cseo}. These evaluations fix the set of sources and measure the answer step. Our
results complement them in two ways: we test the commonly recommended page practices observationally,
within websites, on production engines; and we show that most of the variation in citations arises before
the answer, in retrieval, which a fixed source set excludes. Our observational comparison also relates to work
on search engine optimization that compares optimized with non-optimized pages~\cite{schultheiss2022seo};
we find that the engines' retrieval is not a proxy for classic search rankings.

\paragraph{Retrieval and generation in language models.} Generative search engines are systems of
retrieval-augmented generation~\cite{lewis2020rag} that rewrite the user's input into search
queries~\cite{ma2023rewriting}. Language models use long inputs unevenly~\cite{liu2024lost}, and when they
judge candidates, the order in which the candidates are presented changes the verdict~\cite{wang2024fair}. They also prefer well-known brands:
global over local brands~\cite{kamruzzaman2024brand}, and incumbents over equivalent
products~\cite{chu2026incumbent}. Our controlled experiments measure these effects on the models behind two
production engines, with real questions and sources, and relate them to what the replay shows: the
position effect inflates ChatGPT's preference for its top-ranked sources, and brand names reorder its
rankings in line with what the model believes about them.

\section{Data and Study Design}\label{sec:data}

In this section, we describe the collection, the citations and pages extracted from it, and the design
that separates the effect of a page from that of its website.

\subsection{Answer Collection}\label{sec:collection}

The answers were collected by an academic monitoring platform for generative search engines, operated by
the authors (not named here for anonymity). It sends each query once a day to the public API of each
engine, with web search enabled, no system prompt and default sampling, and stores the answer and its
links. The engines and models are ChatGPT (GPT-5.4 nano with the web search tool), Gemini (Gemini 3.1
Flash Lite with Google Search grounding) and Perplexity (Sonar). The main dataset consists of 651 queries
written for a US fiber internet provider in seven topics (competitive landscape, customer support,
equipment, performance, pricing, reliability, speed), collected daily from 24 June to 21 September 2026:
143,249 answers. Gemini answered only part of the queries until 25 July, so comparisons across engines use
the 26,069 query-days answered by all three. A daily health check found no failed collection day. The
second dataset is described in Section~\ref{sec:generalisation}.

\subsection{Citations and Pages}\label{sec:pages}

A citation is a URL linked in the answer, counted once per answer. Gemini links part of its sources
through a redirect service (1.8\% of all citation rows); we resolve the redirects that are still valid, and
expired ones count at the level of the website, which Gemini names in the link title. URLs are normalized
(lower-case host without \texttt{www.}, no tracking parameters, fragment or trailing slash). The collection
contains 396,188 citations to 39,043 URLs on 7,234 websites (Table~\ref{tab:data}). The engines differ
sharply: ChatGPT cites 2.0 distinct pages per answer and 64\% of its citations go to the providers'
own websites, whereas Gemini cites 10.1 pages, 83\% of them on third-party websites (Reddit alone 13\%),
and Perplexity cites nothing in 96\% of its answers.

\begin{table}[t]
\caption{The main dataset per engine (24 June -- 21 September 2026, 651 queries). \textsuperscript{a}A URL counts once per answer.}
\label{tab:data}
\small
\setlength{\tabcolsep}{3pt}
\begin{tabular}{@{}lrrr@{}}
\toprule
 & ChatGPT & Gemini & Perplexity \\
\midrule
Answers & 58,590 & 26,069 & 58,590 \\
Citations\textsuperscript{a} & 116,530 & 263,300 & 16,358 \\
Pages per answer & 2.0 & 10.1 & 0.3 \\
Answers without a citation & 30.8\% & 15.6\% & 95.8\% \\
URLs / websites & 15,833 / 2,190 & 23,640 / 5,452 & 3,498 / 1,067 \\
Citations to brand websites & 64.3\% & 16.8\% & 6.8\% \\
\bottomrule
\end{tabular}
\end{table}

Every website is labelled as the client's own, a competitor's (the seven other tracked providers and
Verizon, with subsidiaries, parent companies and dealers), or third party; the labels were proposed by
rules and reviewed by hand. Every cited URL was downloaded once by a crawler that respects
\texttt{robots.txt} and waits at least one second between requests to a host; 63.2\% of the 39,059 URLs
returned usable HTML, whose main text we extract with Trafilatura~\cite{barbaresi2021trafilatura}. We
compute fourteen page features (Table~\ref{tab:effects}; definitions in Appendix~\ref{app:data}). Four
measure how well the page matches the queries: word overlap is the Jaccard similarity between the content
words of the page and of all 651 queries, and title, first-paragraph and best-paragraph match are the
highest cosine similarity between that text and any query (all-MiniLM-L6-v2~\cite{reimers2019sbert}).
They compare a page with the whole query set, not with the queries that cited it, so the outcome does not
leak into them. Five features describe structure (length, outbound links, headings, tables, lists), four
mark the practices most often recommended for AI visibility (FAQ block, summary box, author, schema.org
markup), and one is page age. Page speed exists for only 466 pages and is used in Section~\ref{sec:simpson}.
The outcome of a page is $y = \log_2(1 + c)$, where $c$ is its number of citations over the 90 days; the
models use the 12,038 pages cited at least twice, with features, on the 1,120 websites with at least two
such pages.

\subsection{Within-Website Design}\label{sec:design}

A page on a popular website is cited more often than one on an obscure website for reasons unrelated to
the page, so comparing pages across websites attributes website effects to page features. The key idea of
our design is to compare a page only with pages of the same website. For page $p$ on website $w$,
\begin{equation}
y_{pw} = \alpha_w + \mathbf{x}_{pw}^{\top}\boldsymbol{\beta} + \varepsilon_{pw},
\label{eq:fe}
\end{equation}
where $\alpha_w$ is an intercept per website and $\mathbf{x}_{pw}$ contains the page features, indicators
for missing values and the page type. We estimate Equation~\ref{eq:fe} by removing each website's mean,
with robust standard errors~\cite{mackinnon1985hc}. Continuous features are
standardized, so an effect is the percentage more citations per standard deviation (SD), $100(2^{\beta}-1)$;
for the yes/no markers it compares a page that has the marker with one that does not. We fit three samples:
all pages, brand pages (own and competitor websites; 4,794 pages on 77 websites) and third-party pages
(7,185 pages on 1,044 websites).

A single significant coefficient is weak evidence when fourteen features are tested in three samples. We
therefore call a feature \emph{real} only if it passes five tests, each guarding against a different
failure: (1) false-discovery-rate control over the fourteen features
(Benjamini--Hochberg $q<0.05$~\cite{benjamini1995fdr}); (2) selection in at least 60\% of 200 bootstrap
Lasso fits~\cite{tibshirani1996lasso} at the one-standard-error penalty (stability
selection~\cite{meinshausen2010stability}); (3) a double machine learning estimate~\cite{chernozhukov2018dml}
in which all other variables enter through gradient-boosted trees~\cite{ke2017lightgbm}; (4) a monotone or
single-turn partial-effect curve in a generalized additive model~\cite{hastie1990gam}; and (5) the same sign
in at least four of five variants (pages cited at least once or five times, counts capped at the top 1\%,
the first 75\% and the last 25\% of the days). The design removes the website but not unmeasured
differences between pages of a website; the results are associations.

\section{RQ1: Which Page Properties Go With Citations?}\label{sec:rq1}

In this section, we present the results for RQ1, focusing on two aspects: which page properties go with
more citations once the website is held fixed, and whether the practices most often recommended for AI
visibility survive this comparison.

\subsection{Effects Within Websites}\label{sec:within}

Table~\ref{tab:effects} reports the effect of each feature in the three samples, with the number of the
five tests it passes. The main observations are as follows.

First, only the features that measure how well a page matches the queries have robust effects: title and
best-paragraph match pass all five tests, and word overlap passes four (selected in 58\% of the bootstrap
fits, just under the 60\% threshold) and all five on third-party pages. Comparing two pages of the same
website, one SD better best-paragraph match goes with 15.5\% more citations (95\% CI 11.1--20.1\%), title
match with 14.1\% (10.7--17.6\%) and word overlap with 9.1\% (4.4--13.9\%); on third-party pages the effects
are 25.7\%, 23.6\% and 12.8\%. The reason is that the engines retrieve pages by their relevance to the
searches they issue (Section~\ref{sec:rq3}).

Second, no feature passes all five tests on brand pages, whose 77 websites build their pages from shared
templates, so the features vary less within a website. Third, length, outbound links, headings, lists and
tables show no effect, and their partial-effect curves are flat. Page age goes with fewer citations
($-$5.1\% per SD) but is not selected consistently. Fourth, the best-paragraph effect holds for each engine
(ChatGPT +9.5\%, Gemini +7.8\%, Perplexity +4.8\%), whereas title match goes with more citations from Gemini
(+16.8\%) and Perplexity (+4.7\%) but not from ChatGPT ($-$0.2\%), which rewrites a question into its own
searches (Section~\ref{sec:search}).

\begin{table}[t]
\caption{Effect of each page feature on citations, comparing pages of the same website (\% more citations
per SD; yes/no features: page has the marker vs.\ not). In parentheses, the number of the five tests
passed; bold marks features that pass all five.}
\label{tab:effects}
\small
\setlength{\tabcolsep}{3.5pt}
\begin{tabular}{@{}lrrr@{}}
\toprule
Feature & All pages & Brand & Third party \\
\midrule
\MainTabEffectsRows
\bottomrule
\end{tabular}
\end{table}

Links to the page behave differently from the other properties. Using the number of websites linking to
each page in a commercial link index (DataForSEO), one SD more referring websites goes with 22.7\% more
citations within websites, and any referring website with 39.4\% more, mostly for Gemini (+33.7\% per SD,
against +3.7\% for ChatGPT). As the links were measured after the citations, the direction is not
established.

\subsection{Recommended Practices: Across vs.\ Within Websites}\label{sec:simpson}

\begin{figure*}[t]
\centering
\includegraphics[width=0.86\textwidth]{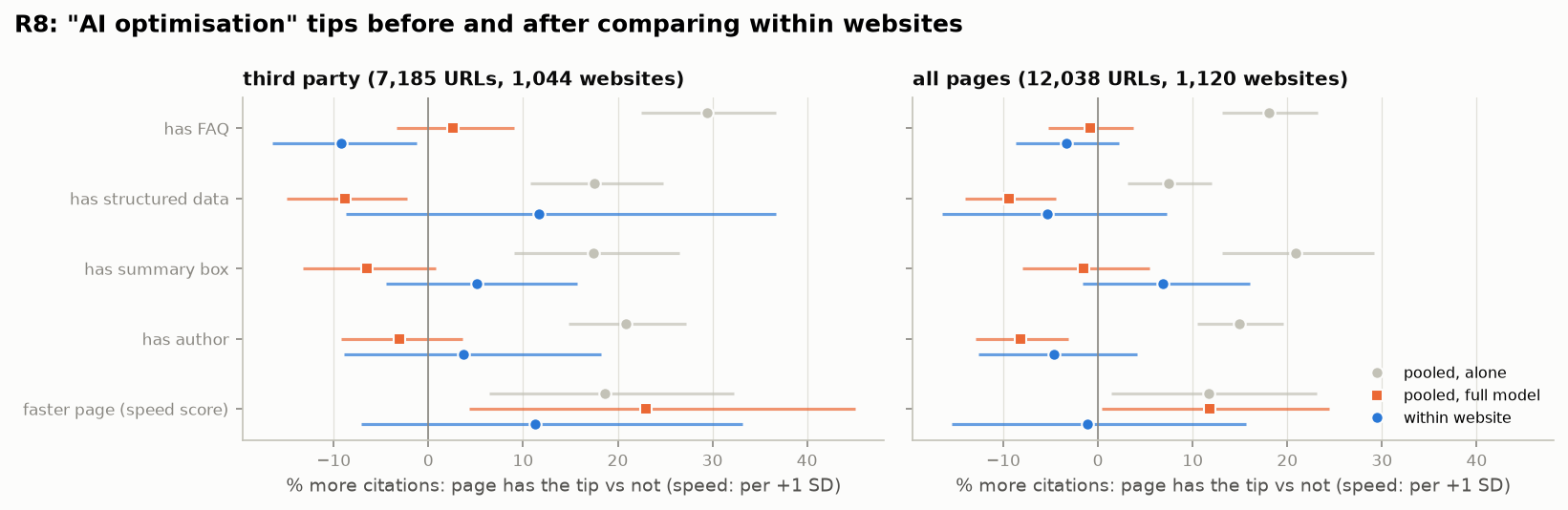}
\caption{Effect of recommended practices on citations: across all pages with the practice alone (grey),
across all pages with all other features (orange), and comparing pages of the same website (blue), with
95\% intervals. Left: third-party pages; right: all pages.}
\Description{Dot plot of the estimated effect of FAQ blocks, structured data, summary boxes, author boxes and page speed under three comparisons; the effects shrink to near zero within websites.}
\label{fig:simpson}
\end{figure*}

Figure~\ref{fig:simpson} compares three estimates for the four recommended markers and page speed: across
websites with the marker alone, across websites with all features, and within websites. Across websites
every practice appears to help: an FAQ block goes with 18.1\% more citations, a summary box 20.9\%, an author
14.9\%, schema.org markup 7.5\% and one SD faster page speed 11.7\%, all significant. Within websites none is
significant ($-$3.4\%, +6.9\%, $-$4.6\%, $-$5.3\%, $-$1.1\%). The reason is that the practices are concentrated
on websites that are cited more for other reasons: FAQ blocks appear on 43\% of the pages of the most-cited
quarter of websites and on 23\% of the least-cited quarter. The across-website comparison credits the
markers with the effect of the website, an instance of Simpson's paradox~\cite{simpson1951}; the same holds
on third-party pages and in the second dataset.

\subsection{Which Queries a Page Is Cited For}\label{sec:perquery}

The match features compare a page with all 651 queries at once. Comparing, within each page, the queries
it is cited for with those it is not, a cited query matches the page better in 88\% of comparisons
(within-page AUC 0.878; all engines 0.87--0.89), and one SD better match raises the odds that the page is
cited for a query by $\times$1.42 (embeddings), $\times$1.44 (title) and $\times$1.68
(BM25~\cite{robertson2009bm25}). Matching also separates cited from never-cited pages: comparing the 4,513
cited pages of the providers' websites with 5,262 never-cited pages from their sitemaps, one SD better
best-paragraph match raises the odds of being cited at all by $\times$11.9. Matching thus decides
\emph{which} queries a page is cited for, but not \emph{how often}: the number of queries a page matches
strongly raises the within-website variation explained only from 4.9\% to 5.9\%. Section~\ref{sec:rq2}
examines the remaining variation, and Section~\ref{sec:rq3} shows where content acts.

\section{RQ2: How Much Do Page Properties Explain?}\label{sec:rq2}

In this section, we measure how much of the variation in citations the website and the page features
explain, and we characterize the rest. We focus on two aspects: whether the remainder is chance, and
whether it is a fixed property of pages or changes over time.

\subsection{Website, Page and Chance}\label{sec:variance}

In a mixed model with a random intercept per website, 13\% of the variation in citation counts lies between
websites and 87\% between pages of the same website. The page features explain 42\% of the between-website
part but only 3.4\% of the within-website part. To measure chance, we count each page's citations
separately on odd and even days: the two counts correlate with $r=0.82$, a reliability of 0.90 for the
90-day count (Spearman--Brown; 0.89 within websites). Thus only about 10\% of the variation is day-to-day
chance, and about 75\% is repeatable, lies within websites and is explained by no page feature. Nor does the
website predict unseen pages well: on held-out pages it explains 4.0\% of the variation (11.4\% for websites
with at least 30 pages).

\subsection{Breadth, Depth and the Page--Query Bond}\label{sec:bond}

A page can be cited often because it is cited for many queries, or because it is picked repeatedly for a
few. We split the outcome into breadth, $B=\log_2(1+b)$ for the number $b$ of queries the page is cited for,
and depth, $D=y-B$; they account for 47\% and 53\% of the variance. In every engine, page features explain
4--12\% of the within-website variation in breadth but only 1--4\% in depth, so they speak to which queries a
page is cited for and almost not to how often it is picked for them. To study depth directly, we treat
every answer as a choice among the pages that compete for its query (300 random queries; candidates are the
pages cited for the query in the first 60 days, outcomes the answers of the last 30 days, about 5,000 per
engine). In a conditional logit, the page's match to the query explains 0.6\% (ChatGPT) and 2.8\% (Gemini) of
the choice (McFadden $R^2$), its citations for other queries raise this to 5.5\% and 5.8\%, and its own
history with this query to 19.2\% and 25.6\%. Depth is thus a persistent bond between a page and a query
that neither the page's match nor its general quality captures; Section~\ref{sec:rq3} shows that it is the
page's position in the engine's own search for the query.

\subsection{Drift}\label{sec:drift}

Odd and even days both span the whole 90 days, so a page cited heavily in July and not at all in September
counts as repeatable. To measure stability over time, we split the collection into six 14-day windows and
compute the share of within-website variation that two windows have in common. Within a window, 63--67\% is
repeatable. Two weeks later, 40\% (ChatGPT), 47\% (Gemini) and 22\% (Perplexity) is still shared, and ten
weeks later only 14\%, 29\% and 0\%. The cited set changes steadily, because the engines regenerate their
searches for every answer and their search results change over weeks (Section~\ref{sec:rq3}). This limits any
explanation built on a single snapshot of page features: a large part of the 75\% is not a fixed property of
pages at all.

\section{RQ3: Retrieval and Model Knowledge}\label{sec:rq3}

In this section, we open the step between the question and the answer: the searches the engine runs, the
pages it retrieves and what the model already knows. We focus on four aspects: how closely citation
follows the engine's own retrieval, how the engines search, how much of the variation retrieval explains,
and what the model contributes.

\subsection{Replaying the Engines}\label{sec:replay}

The collection records only the answer and its citations. To observe retrieval, we send the 651 questions
to the same models through their APIs with web search enabled (gpt-5.4-nano with the \texttt{web\_search}
tool; gemini-3.1-flash-lite with Google Search grounding), \gfReplayRuns{} times on \gfReplayDays{} days up
to \gfReplayLastDay{}. For ChatGPT, the API returns the queries of every search call and the list of sources
it returned; one call sends about four queries and returns a single pooled, re-ranked list, and the rank
of a page is its position in that list. Gemini's grounding metadata gives the queries and the sources used,
not a ranked list, so the rank analyses use ChatGPT; Perplexity has no replay. The replay behaves like the
collection: when it did not search, the collected answers to the same question had no citation in
\gfZeroIfNotSearched\% of the cases (\gfZeroIfSearched\% when it did), and \gfReplayInCollected\% of the pages
it cites were cited in the collection's last 14 days, about as many as the collection repeats itself
(\gfCollectedDayToDay\%). Every page ChatGPT cites is one its searches returned (\gfCitedRetrievedPct\% of
\gfCitedURLs{} cited URLs; \woCitedRetrievedPct\% of \woCitedURLs{} in the second dataset): a page is cited
only after it has been retrieved.

\subsection{Retrieval and Selection}\label{sec:selection}

For each question, ChatGPT consults \gfConsultedPerQ{} pages on average and cites \gfCitedPerQ{}, following
its own ranking: over \gfSourceLists{} source lists, it cites \gfRankOne\% (95\% CI
\gfRankOneLo--\gfRankOneHi\%) of the pages it ranks first, \gfRankSixTen\% at ranks 6--10 and
\gfRankElevenTwenty\% at ranks 11--20, and a page returned by several calls has \gfMultiCallOR{} times the odds
of being cited. To separate retrieval from selection, we download the consulted pages and fit a conditional
logit to the choice within each list. The rank alone explains \gfSelRank\% of the choice (McFadden $R^2$);
adding the match to the question and to the engine's own queries and all page features raises this only to
\gfSelPage\%. Content still matters at equal rank, in a consistent direction: brand pages are cited more
($\times$\gfSelBrand), as are pages with more headings ($\times$\gfSelHeadings) and shorter pages
($\times$\gfSelLength{} per SD of length), whereas lists ($\times$\gfSelList), comparisons
($\times$\gfSelComparison) and Reddit threads ($\times$\gfSelReddit) are consulted often but cited rarely.
The recommended practices do not help at this step either: at equal rank, author boxes ($\times$0.77), FAQ
blocks ($\times$0.93) and schema.org markup ($\times$0.92) go with slightly lower odds of being cited. Content
thus matters mainly through the ranking: what counts is which pages the engine retrieves and how high.

\subsection{How the Engines Search}\label{sec:search}

ChatGPT splits a question into about \gfGptQueriesPerAnswer{} keyword-style queries, keeps
\gfGptWordsKept\% of the question's words, adds the current year to \gfGptAddYear\% of its queries and a brand
the user did not name to \gfGptAddBrand\%, and often searches again for vendors found in the first round.
Gemini writes about \gfGemQueriesPerAnswer{} natural-language sub-questions closer to the question (cosine
similarity \gfGemSimilarity{} against \gfGptSimilarity), rarely adding a year or a brand (\gfGemAddYear\% and
\gfGemAddBrand\%). Both rewrite the question anew on every answer: between two runs on the same day,
ChatGPT's queries overlap by \gfSameQueries\% (Jaccard) and its cited pages by \gfSameCited\%. Although
ChatGPT's queries look like keyword queries, it ranks pages by meaning: the rank correlates more with a
page's best embedding similarity to the queries ($\rho=$\gfRhoMeaning) than with its best BM25 score
($\rho=$\gfRhoKeywords), and with both included the match in meaning has the larger effect (+0.22 against
+0.05 per SD). Text match explains little of the order, however; the ranking reflects signals we do not
observe.

Nor is the ranking that of the public search engines. For the question as asked, 0.1--0.6\% of the cited
pages are in Google's top 10; for the engines' own queries, 0.5\% (Bing) and 0.7\% (Google) of the pages
ChatGPT retrieves and 5.6\% of those Gemini cites are in the top 10, although Gemini is grounded on Google
Search. The pages are not further down either: \gfDeepGptRetGoogleAbsent\% of ChatGPT's retrieved pages are
absent from Google's top 50 and \gfDeepGptRetBingAbsent\% from Bing's top 30 for all of the answer's queries,
as are \gfDeepGemCitGoogleAbsent\% of Gemini's cited pages, whereas the websites of half of ChatGPT's pages
(\gfDeepGptRetGoogleSiteFound\%) do appear in Google's top 50: the engines select different pages, often from
the same websites. The public results were retrieved with a US location within one to two days of the
replay, so time and place cannot account for the gap. One outside signal does enter: links to the page
(Section~\ref{sec:within}), mostly for Gemini, act on which pages are retrieved but not on which retrieved
pages ChatGPT cites (odds ratios 0.97--1.07, not significant).

\subsection{How Much Retrieval Explains}\label{sec:explains}

We now return to the outcome of Section~\ref{sec:rq2}, summarizing the replay per page by how often it was
retrieved and how often it was among the top five of a list. The results are presented in
Table~\ref{tab:retrieval}. Retrieval explains \gfRetrieval\% (95\% CI \gfRetrievalLo--\gfRetrievalHi\%) of the
within-website variation in ChatGPT's 90-day citation counts (\gfVarPages{} pages, \gfVarSites{} websites),
against \gfPageVar\% for all page features, a difference of \gfRetrievalMinusPage{} points (CI
\gfRetrievalMinusPageLo--\gfRetrievalMinusPageHi); shuffling the replay variables between pages of the same
website removes the effect (\gfShuffled\%, $p=$\gfShuffledP). Consistent with the drift, the replay, which
took place after the collection, explains \gfWindowFirst\% of the citations of the first two weeks and
\gfWindowLast\% of the last days. The page--query bond of Section~\ref{sec:bond} is the page's place in the
engine's retrieval for the query.

\begin{table}[t]
\caption{What the replay shows (ChatGPT; 95\% CI in brackets).}
\label{tab:retrieval}
\small
\begin{tabular}{@{}p{5.2cm}r@{}}
\toprule
Pages consulted / cited per question & \gfConsultedPerQ{} / \gfCitedPerQ \\
Cited at rank 1 / 2 / 6--10 / 11--20 (\%) & \gfRankOne{} / \gfRankTwo{} / \gfRankSixTen{} / \gfRankElevenTwenty \\
Choice among consulted pages explained: rank / + content & \gfSelRank\% / \gfSelPage\% \\
Same question, same day: query / cited-page overlap & \gfSameQueries\% / \gfSameCited\% \\
Retrieved pages absent from Google top 50 / Bing top 30 & \gfDeepGptRetGoogleAbsent\% / \gfDeepGptRetBingAbsent\% \\
Within-website variation explained: retrieval & \gfRetrieval\% [\gfRetrievalLo, \gfRetrievalHi] \\
\quad page features / shuffled retrieval & \gfPageVar\% / \gfShuffled\% \\
Citations from websites the model names without search & \gfPriorCitations\% (baseline \gfPriorBaseline\%) \\
\bottomrule
\end{tabular}
\end{table}

\subsection{The Model's Prior}\label{sec:prior}

To measure what the model knows, we ask each engine's model, without search, which brands and websites it
would recommend for every question. ChatGPT writes a brand into its queries in \gfPriorAddIfIn\% of the cases
when the brand is in its prior for the question, against \gfPriorAddIfOut\% otherwise. Websites in the prior
rank \gfPriorRankFactor{} times higher in its lists and have \gfPriorSelOR{} times the odds of being cited at
equal rank, and \gfPriorCitations\% of ChatGPT's citations come from websites its model names without
searching (\gfPriorBaseline\% with the prior of another question). The models know different markets: for the
second dataset, the model behind ChatGPT names the tracked vendors for 2--16\% of the questions and that
behind Gemini for 60--96\%. The prior also reaches the answer without any page: \gfGptMentionNone\% of
ChatGPT's mentions of a brand the question does not name are supported by no cited page that names it, and a
brand in the prior is named with $\times$\gfGptMentionPriorOR{} the odds at equal citation support. Pages are
cited only after they are retrieved, whereas brands can be named from memory.

\subsection{From Sources to Answer}\label{sec:answerstep}

The replay shows which pages reach the model, not how the model chooses among them. We complement it with
controlled experiments on the answer step, in which the input is fixed and one factor is varied at a time:
the position of a source, the names of brands and publishers, and whether the model's preferences can be
described by one score per brand. The experiments use the models behind the replayed engines without
search, real questions from both datasets, and real sources from the replay or fixed product profiles
written from the brands' own pages (Appendix~\ref{app:controlled}). The results are presented in
Table~\ref{tab:answerstep}. The main observations are as follows.

First, ChatGPT's model cites what it reads first, whereas Gemini's does not. Showing 8 real sources for each
of 100 questions in all 8 orders (9,600 answers), the same source is cited in 60\% of ChatGPT's answers when
shown first and 43\% when shown last (17.6 points, 95\% CI 15.1--20.3), in line with the position bias of
language models~\cite{liu2024lost}; Gemini cites it in 57--59\% in every position. The engine's own ranking
still predicts citation when the position is randomised, so part of ChatGPT's rank curve
(Section~\ref{sec:selection}) is a position effect, and none of Gemini's.

Second, brand names change the ranking of products with identical facts; publisher names hardly do. In 60
packets, each with a real question, five products described by profiles of equal length and one
third-party excerpt per product, we reveal or mask brand and publisher names independently (1,920
answers). Since every answer places exactly three products in its top three, revealing names cannot raise
all brands at once; we therefore compare the change in each product's top-three rate with that between two
identical prompts. Revealing brand names changes the rates 2.1 (ChatGPT) and 3.3 (Gemini) times as much as
repetition, revealing publisher names 1.3 and 1.6 times. Spectrum loses 33 points (ChatGPT) once named,
whereas Google Fiber gains 12 and the niche vendor Loopio 16. When the prompt states that the names were
assigned at random and carry no information, the same profile still reaches ChatGPT's top three in 80\% of
the answers under one name and 39\% under another, and a brand's gain when revealed follows this name
effect (Spearman 0.81), not whether its masked profile can be identified. The name brings in what the model
believes about the brand; a cable provider is ranked lower for fiber questions on identical facts.

Third, within a question the model's preferences are described by one score per brand; across questions they
are not. In the collected answers, the brand of a pair that is named first is the same in 78--81\% of the
answers to a question, majority cycles are rare (one of 760 questions after resampling), and a
Bradley--Terry model~\cite{bradley1952} with one score per brand and question predicts 74--79\% of the pair
orders on held-out days, against 64--67\% for one score per brand overall. With the evidence fixed (20
questions, six brands, 6,000 answers), one score per brand fitted on pairwise answers orders 82\% (ChatGPT)
and 77\% (Gemini) of the pairs in full rankings correctly, about as well as one full ranking predicts another
shown in a different order (81\% and 79\%). Removing or adding a product, or rewording the question, changes
the ranking little more than repetition; the order in which products are listed changes it most. The same
structure holds when the model searches the web for its own evidence (3,480 answers), but consistency then
depends on the evidence: two identical prompts agree with Kendall $\tau$ 0.75 (ChatGPT) on fixed profiles,
0.56 with search and 0.51 from memory alone. For the niche vendors, which the models barely know, answers
from memory are the least consistent ($\tau$ 0.36) and search restores consistency (0.59).

We stress that these experiments measure the small models behind the engines on constructed inputs. They
show that the answer step adds two effects to retrieval: the order in which the engine presents its sources,
and what the model believes about the brands they describe.

\begin{table}[t]
\caption{Controlled experiments on the answer step (models without search unless stated; 95\% CI in
brackets) and brand order in the collected answers.}
\label{tab:answerstep}
\small
\begin{tabular}{@{}p{4.1cm}rr@{}}
\toprule
 & ChatGPT & Gemini \\
\midrule
Same source first $-$ last, cited (points) & 17.6 [15.1, 20.3] & $-$0.4 [$-$2.3, 1.6] \\
Brand names shown: change vs repeat & 2.1$\times$ [1.8, 2.6] & 3.3$\times$ [2.5, 4.5] \\
Publisher names shown: change vs repeat & 1.3$\times$ [1.1, 1.6] & 1.6$\times$ [1.3, 2.0] \\
Random names: spread in top-3 rate (points) & 41 & 18 \\
Held-out pairs, collected answers: score per brand and question / per brand & 74 / 64\% & 77 / 67\% \\
Full rankings predicted: one score per brand / other order & 82 / 81\% & 77 / 79\% \\
Kendall $\tau$ of repeats: fixed profiles / search / memory & .75 / .56 / .51 & .89 / .76 / .82 \\
\bottomrule
\end{tabular}
\end{table}

\section{Generalisation: A B2B Niche Market}\label{sec:generalisation}

The main dataset describes a consumer market dominated by a few large brands. In this section, we repeat
the core analyses on a market of the opposite kind, many small and little-known vendors of business
software, and ask which findings hold in both and which depend on the market.

The second dataset consists of 55 queries about proposal and business-development software for
architecture, engineering and construction (AEC) firms, written for a small vendor in that market and
collected by the same platform, from the same engines, with the same settings and over the same 90 days:
14,850 answers with 71,849 citations to 9,258 URLs on 3,895 websites. Gemini accounts for 84\% of the
citations, ChatGPT cites something in only 35\% of its answers (69\% in the main dataset), and vendor
websites receive about half of every engine's citations. Website roles were proposed by a language model
for every website with at least ten citations and reviewed by hand. The within-website models rest on
2,477 pages on 515 websites, a quarter of the main dataset.

Table~\ref{tab:general} compares the two markets. The main observations are as follows. First, the
findings about content replicate: title and best-paragraph match pass all five tests (title
\woTitleAll\% per SD, \gfTitleAll\% in the main dataset), now also on brand pages (\woTitleBrand\%), and the
recommended practices look effective across websites but not within them (FAQ block \woFaqPooled\% vs.\
\woFaqWithin\%). Second, the website matters much less: only \woBetween\% of the variation lies between
websites (\gfBetween\%) and website authority has no effect (\woSiteAuthority\% per SD, \gfSiteAuthority\%),
as small vendors have no authority advantage to lean on. Third, the retrieval findings replicate: ChatGPT
cites \woRankOne\% of the pages it ranks first and \woRankElevenTwenty\% of those at ranks 11--20, rewrites
every question anew, and \woDeepGptRetGoogleAbsent\% of its retrieved pages are absent from Google's top 50.
Retrieval explains \woRetrieval\% (CI \woRetrievalLo--\woRetrievalHi\%) of the within-website variation
against \woPageVar\% for page features, although with \woVarPages{} pages the difference is not significant
(CI \woRetrievalMinusPageLo{} to \woRetrievalMinusPageHi). Fourth, the models' knowledge differs by market.
The model behind ChatGPT names the tracked vendors for 2--16\% of the queries (61--72\% for the large
providers), and the one behind Gemini for 60--96\%. ChatGPT therefore rarely writes brands into its queries
(\woGptAddBrand\% against \gfGptAddBrand\%) and instead learns vendor names from a first round of results, and
a cited third-party page that names a vendor raises the odds that the answer names it
$\times$\woGptMentionThirdOR{} (ChatGPT) and $\times$\woGemMentionThirdOR{} (Gemini), against
$\times$\gfGptMentionThirdOR{} and $\times$\gfGemMentionThirdOR{} in the main dataset.

\begin{table}[t]
\caption{Core results in the two markets. Page effects are within websites (\% per SD; yes/no markers: has
vs.\ not; in parentheses, tests passed). Replay results for ChatGPT.}
\label{tab:general}
\small
\setlength{\tabcolsep}{3pt}
\begin{tabular}{@{}p{3.2cm}rr@{}}
\toprule
 & Consumer & Niche B2B \\
\midrule
Title matches a query, all / brand pages & \gfTitleAll{} (\gfTitleAllTests) / \gfTitleBrand{} (\gfTitleBrandTests) & \woTitleAll{} (\woTitleAllTests) / \woTitleBrand{} (\woTitleBrandTests) \\
Best paragraph matches, all pages & \gfBestParaAll{} (\gfBestParaAllTests) & \woBestParaAll{} (\woBestParaAllTests) \\
FAQ block: across / within websites & \gfFaqPooled{} / \gfFaqWithin & \woFaqPooled{} / \woFaqWithin \\
Variation between websites & \gfBetween\% & \woBetween\% \\
Website authority / page links (per SD) & \gfSiteAuthority{} / \gfPageLinks\% & \woSiteAuthority{} / \woPageLinks\% \\
Rank 1 / ranks 11--20 cited & \gfRankOne{} / \gfRankElevenTwenty\% & \woRankOne{} / \woRankElevenTwenty\% \\
Retrieval vs.\ page features & \gfRetrieval{} vs.\ \gfPageVar\% & \woRetrieval{} vs.\ \woPageVar\% \\
Citations from the model's prior websites & \gfPriorCitations\% & \woPriorCitations\% \\
\bottomrule
\end{tabular}
\end{table}

In summary, the content findings, the dominance of the engine's own ranking and the rewriting of queries
hold in both markets. What depends on the market is the weight of the website and of the model's prior
knowledge: both matter for large consumer brands, whereas small vendors depend on being found by search and
on third-party pages that name them.

\section{Discussion}\label{sec:discussion}

\paragraph{Implications for publishers.} Four conclusions follow. First, write pages for the questions
people ask: a title and a paragraph that answer a question in its own terms are the only page properties
that pass every test, most strongly on third-party pages (\gfTitleThird\% and \gfBestParaThird\% per SD), and
they decide which questions a page is cited for. Second, being retrieved matters more than anything on the
page. The engines search with their own rewritten queries, rank pages by how well they match these queries
in meaning, and do not return Google's or Bing's pages, so a good rank in classic search is not a proxy; at
equal rank, concise pages from the original source win over roundups and forums. Third, the practices most
often recommended for AI visibility (FAQ blocks, schema.org markup, summary and author boxes, page speed) are
not associated with more citations within a website, and at equal rank they go with slightly fewer. Fourth,
citations must be monitored over time and per question: the cited set changes over weeks and a brand's
standing is specific to the question. Replaying the questions through the engines' APIs, as we do, shows
which pages are retrieved, at what rank, and which are retrieved but not cited.

\paragraph{Known and unknown brands.} The results point to two routes to visibility. For the large
providers of the main dataset, the ranking largely resides in the model's memory: without any evidence,
the models rank them consistently (one score per brand predicts 75\% of the pairs for ChatGPT and 92\% for
Gemini), and search loosens the ranking slightly (76\% and 89\%). For the small vendors of the second
dataset, which the models barely know, there is no stable ranking without evidence (64\% and 67\%), and
search restores it to about 80\%, close to the level with fixed facts. This matches the brand-level results:
a known brand is named even without a supporting page and gains when its name is revealed, whereas a niche
vendor is named mainly when a cited third-party page names it ($\times$\woGptMentionThirdOR{} for ChatGPT,
against $\times$\gfGptMentionThirdOR{} in the main dataset). For an established brand, visibility is largely
set by what the model already believes; for a niche brand, a consistent position exists only when the
engine's search finds evidence about it.

\paragraph{A hypothesis: the prior as an entry point to retrieval.} Our results trace a chain from the
model's prior to citation: a brand in the prior is written into the engine's searches (\gfPriorAddIfIn\%
against \gfPriorAddIfOut\%), these searches retrieve the brand's pages and third-party pages about it, which
are preferred at equal rank (websites in the prior $\times$\gfPriorSelOR, brand pages $\times$\gfSelBrand),
and a cited page about the brand leads to the brand being named; the prior also acts without any page. We
therefore hypothesize that, if a brand's representation in a model's prior could be changed, an effective
strategy would combine placing accurate associations between the brand and the relevant questions into the
prior with making the brand's own and third-party pages retrievable for the brand-named queries this
produces. Two conditions follow from our data: what is imprinted matters, not only the name (a name can lower
a brand's ranking on identical facts); and the prior alone yields mentions without citations, since every
cited page is retrieved. We stress that this is a hypothesis: our study establishes associations along the
chain, not the effect of changing a prior, and known brands also differ in size, coverage and links.
Testing it requires intervention studies, for example probing priors again after model updates.

\paragraph{Implications for research.} Studies of generative engine optimization have mostly evaluated
content changes on a fixed set of retrieved sources~\cite{aggarwal2024geo}. Such evaluations measure only
the answer step, whereas most of the variation we observe arises in retrieval; they should also randomize
the order of the sources and report the brands involved, since position and names change the answer on
identical content. Observational comparisons should hold the website fixed, as every recommended practice
appears effective across websites and none within them, and should be longitudinal, as citations drift.

\paragraph{Limitations.} The results are associations: the within-website design removes everything constant
within a website, but not unmeasured differences between its pages, and page links were measured after the
citations. The pages were downloaded once, after the collection, and about a third of the cited URLs could
not be analysed. The replay uses the engines' APIs after the collection, which need not behave like their
consumer applications, and Perplexity could not be replayed. Both datasets consist of English questions
about the US market, written for the organizations that use our monitoring platform rather than sampled
from user logs. The controlled experiments use the small models behind the engines and constructed inputs,
and the second dataset is small. Finally, the engines change continuously; our measurements describe them
in 2026, and the released replay logs allow later studies to test whether the mechanisms persist.

\section{Conclusion}\label{sec:conclusion}

We studied what shapes citation in generative search engines, using 90 days of answers in two contrasting
markets, an API replay of the engines' retrieval and controlled experiments on the answer step. Content
matters in one way: pages whose title and paragraphs match the questions people ask are cited more, whereas
the commonly recommended practices are not associated with more citations within a website. Page properties
explain little of the repeatable variation between pages; the engine's own retrieval explains much more, and
every page ChatGPT cites is one its searches returned, ranked in an order unlike that of Google or Bing. The
model's knowledge enters every step: it shapes the searches, the preference among retrieved pages, and which
brands are named, with or without a supporting page. For publishers, being found by the engine's own search
matters more than any page feature; for researchers, evaluations must cover retrieval and hold the website
fixed. We release the data, replay logs and code to support such studies.

\section*{Ethics Statement}

The study analyses answers of public generative search engines to commercial questions and publicly
accessible Web pages; it involves no human subjects and no personal data beyond what appears on public pages.
The engines were queried through their official APIs within their terms of use and rate limits. The page
crawler respected robots.txt and rate limits but used a browser user agent and did not identify itself as a
research crawler; we did not fetch pages from websites that disallow automated access, such as Reddit. We do
not redistribute the downloaded pages. Brand names appear because the questions concern commercial markets;
our results describe the engines' behaviour, not the quality of any brand.

\section{Acknowledgments}

We wish to thank the team at ktau.ai for their assistance with the data in our analysis.

\bibliographystyle{ACM-Reference-Format}
\bibliography{references}

\appendix

\section{Data and Features}\label{app:data}

\paragraph{Questions.} Table~\ref{tab:app-questions} lists the questions per topic and intent. Intents were
labelled by a language model (gpt-5.4-mini) with keyword rules as a fallback, into \emph{learning}
(general questions), \emph{comparing} (brands or options) and \emph{ready to buy} (prices, fees, contracts,
switching). The second dataset consists almost entirely of learning questions about software capabilities;
its three comparison questions are too few to analyse intent separately.

\begin{table}[h]
\caption{Questions per topic and intent.}
\label{tab:app-questions}
\small
\AppTabQuestions
\end{table}

\paragraph{Website roles.} In the main dataset, 16 domains belong to the client, 217 to the competitors
(including subsidiaries, parent companies and dealer websites) and 7,186 are third party; they receive
11.1\%, 13.8\% and 75.0\% of the citations. In the second dataset, where the competitors are not known in
advance, a language model proposed the owner of every website with at least ten citations from its home
page and the citing answers, and the proposals were reviewed by hand; competitor websites receive 46.7\%
of the citations, the client's 2.6\% and third-party websites 50.8\%. Websites of an unrelated company
with a similar name were moved to third party, and a vendor's rebranded domain was assigned to the vendor.

\paragraph{Download.} Every cited URL was requested once. In the main dataset, 63.2\% of the 39,059 URLs
returned usable HTML, 18.9\% were disallowed by \texttt{robots.txt} (and not requested), 7.3\% refused the
request (HTTP 403), and 3.1\% and 3.0\% were PDF documents and videos. In the second dataset, 76.9\% of
the 9,258 URLs returned usable HTML, 12.6\% refused the request and 1.5\% were disallowed. For the
selection analysis of Section~\ref{sec:selection} only, PDF documents were parsed and Wikipedia pages were
obtained through the Wikipedia REST API.

\paragraph{Features.} The match features are defined in Section~\ref{sec:pages}. The remaining features
are computed on the main text of the page. \emph{Length} is the number of words; \emph{outbound links} the
number of distinct external hosts linked; \emph{headings}, \emph{tables} and \emph{lists} are counts of
the corresponding elements. An \emph{FAQ block} is a heading containing ``FAQ'', ``frequently asked
questions'' or ``common questions'', or schema.org \texttt{FAQPage} markup. A \emph{summary box} is a
heading or an opening passage labelled ``TL;DR'', ``key takeaways'', ``at a glance'', ``quick answer'',
``the short answer'', ``bottom line'' or ``summary''. An \emph{author} is a byline element or an author in
the structured data; forum pages, which show user names, are counted as having no author.
\emph{Schema.org markup} is any JSON-LD or microdata type. \emph{Page age} is the number of days between
the publication or update date of the page and its median citation day; dates after the citation are
extraction errors (the page was downloaded later) and are treated as missing. The \emph{page type} is
assigned by ordered rules on the host, the URL path and the title: video and forum hosts; support and
pricing paths; comparison, review-list, list, how-to and pricing words; shallow product paths; and
editorial paths or at least 600 words for articles.

\paragraph{Hand checks.} For a random sample of 50 pages per dataset, we compared the extracted text,
date, page type and markers, and the role of the page's website, with the page itself: all were correct
for all 50 pages in both datasets (95\% upper bound on the error rate 6\%). A further sample of 195 pages of the main dataset, up to 19 per page
type, confirmed every page-type label. The language model's intent label was correct for all 100 random
questions of the main dataset checked and all 55 questions of the second.

\section{Full Robustness Results}\label{app:robust}

Table~\ref{tab:app-tests} gives the five tests of Section~\ref{sec:design} for the fourteen features on all
pages of both datasets. In the main dataset, title and best-paragraph match pass all five tests. Word overlap,
first-paragraph match and page age fail only the stability selection; word overlap is selected in 58\%
of the bootstrap fits, just under the 60\% threshold, and passes all five tests on third-party pages. The recommended practices pass only the sign test
and the shape test (which does not apply to yes/no markers): they are neither significant after FDR
control nor selected, i.e., they are consistently close to zero. In the second dataset, title and
best-paragraph match pass all five tests and word overlap four; headings pass four tests there but only
two in the main dataset.

\begin{table}[h]
\caption{The five tests on all pages: 1 FDR significance, 2 stability selection, 3 double ML, 4 shape,
5 sign in $\ge$4 of 5 variants ($\bullet$ passed, $\circ$ failed). Effects in \% more citations per SD
($^\dagger$ has vs.\ not). Bold: passes all five.}
\label{tab:app-tests}
\small
\setlength{\tabcolsep}{3pt}
\AppTabTests
\end{table}

Table~\ref{tab:app-engines} repeats the within-website model for the citations of each engine
separately. Best-paragraph match holds for every engine, whereas title match is associated with more
citations from Gemini and Perplexity but not from ChatGPT; the reason is that ChatGPT rewrites the question
into its own searches (Section~\ref{sec:search}). Older pages receive fewer citations from ChatGPT and
Gemini. FAQ blocks and schema.org markup have no effect for any engine.

\begin{table}[h]
\caption{Within-website effects per engine, main dataset (\% per SD; $^\dagger$ has vs.\ not; $^{*}$ $p<0.05$).}
\label{tab:app-engines}
\small
\AppTabEngines
\end{table}

\section{Additional Analyses}\label{app:additional}

\paragraph{Website authority.} Across websites, one SD of any of five authority measures (Common Crawl
harmonic centrality and PageRank, Majestic referring subnets, Tranco and CrUX rank) goes with 21--30\% more
citations. Once the number of cited pages of the website is controlled, the association turns negative
($-$5 to $-$8\%): authoritative websites have more cited pages, not more citations per page. In the
page-level mixed model, one SD of the website's harmonic centrality goes with 6.6\% more citations for a
page, much less than the page's own match to the questions (13.5\% for title match).

\paragraph{Page type.} Within websites and relative to articles, forum pages receive 78\% more citations
(95\% CI 40--127\%) and list pages 13\% more (2--25\%); comparison, how-to, pricing, product and support
pages do not differ significantly from articles. Reddit alone accounts for 2.2\% of ChatGPT's and
13.1\% of Gemini's citations, although at equal rank ChatGPT rarely cites Reddit threads it has
retrieved (Section~\ref{sec:selection}).

\paragraph{Position of the matching paragraph.} The paragraph that best matches the questions lies at a
median of 45\% of the page length; it is in the top quarter for 32\% of the cited pages and in the bottom
quarter for 28\%. It need not be at the top: the engines find it wherever it is.

\paragraph{Entity coverage.} Pages that name more of the tracked brands receive more citations
(+11.8\% per SD within websites), mainly by being cited for more questions (+10.3\% in breadth); the
number of prices and places mentioned has no effect.

\paragraph{Crawler access and JavaScript.} Blocking in \texttt{robots.txt}, read at download time, does not
keep pages out of the answers: 7.1\% of the pages ChatGPT cites disallow its search crawler
(OAI-SearchBot), and 19.9\% of those Gemini cites disallow Googlebot. For 11.5\% of the downloaded pages the
main text requires JavaScript; within websites they receive fewer citations from Gemini ($-$10.6\%,
$p=0.008$) and, not significantly, from ChatGPT ($-$6.8\%) and Perplexity ($-$2.1\%).

\section{Controlled Experiments on the Answer Step}\label{app:controlled}

\paragraph{Common setup.} The experiments call gpt-5.4-nano (ChatGPT) and gemini-3.1-flash-lite (Gemini)
through their APIs, without web search unless stated. The instructions restrict the model to the
information in the message; answers are returned as JSON or with source IDs in brackets. All answers
parsed, except 9\% of Gemini's pairwise answers in the stability experiment, where it declined to choose.
Confidence intervals are bootstrapped over questions or packets. The released repository contains the
full protocols and prompts.

\paragraph{Position.} For 100 questions (80 and 20 from the two datasets), 8 replay sources are shown as
title and the three paragraphs most similar to the question, under random IDs, in all 8 cyclic orders,
without labels, with rank labels aligned with the position and with independent rank labels, twice
(9,600 answers). The position effect is estimated for the same source in different positions (source
fixed effects, errors clustered by question).

\paragraph{Product profiles.} For 19 brands (nine providers; ten vendors), gpt-5.4-mini writes a profile
of seven fields and 150--195 words from the brand's own pages, referring to it by a placeholder; comparison
pages and paragraphs naming other brands are excluded. Automatic checks reject names, products, domains
and executives of any brand of the market and fields above 30 words. In a leak test, the well-known
providers are identified from their facts above chance (ChatGPT 26--33\%, Gemini 52--56\%; chance about
10\%), the vendors never.

\paragraph{Brand and publisher identity.} Sixty packets combine a real question, five brands and, per
brand, one third-party excerpt (35--140 words, naming only that brand) from a well-known (Tranco rank
$\le$ 20,000) or little-known publisher. Brand and publisher names are masked or revealed independently,
with facts, wording, order and length identical; each packet is shown in two orders, twice (1,920
answers). Since every ranking has three products in its top three, we compare the RMS change of a product's
top-three rate with that between identical prompts. In a hypothetical version, the prompt states that
names were assigned at random and rotates them over the profiles (1,200 answers); the name effect is a
brand's top-three rate relative to that of the same profile.

\paragraph{Brand order in collected answers.} Tracked brands not named in the question are ordered by first
mention. Pairs with at least 10 common answers per question give consistency and cycles (a cycle counts if
it persists in 90\% of 200 resamples). Penalized Bradley--Terry models at three levels (engine, topic,
question) are fitted on half of the days and evaluated on the other half.

\paragraph{Controlled stability.} Twenty sets combine a question and six brands described by the profiles
under their real names. For each set the model answers all pairs in both orders three times, ranks all six
in six orders twice, ranks five after each removal, seven after an addition, and all six under three
rewordings (6,000 answers). A Bradley--Terry model fitted on the pairwise answers is evaluated on the pairs
of each full, reduced, extended or reworded ranking. The search version repeats this design in three arms
that differ only in the evidence (profiles, web search, memory), with the same answer format (10,440 answers,
3,480 per arm); the profile arm reproduces the original ($\tau$ 0.75 against 0.77 for ChatGPT).
Regressing the agreement of two searched prompts on the overlap of their sources gives a slope of 1.25
(95\% CI 0.48--2.01) for ChatGPT and 0.26 (0.07--0.46) for Gemini.

\end{document}